\documentclass[
aps,pre,
amsmath,amssymb,
longbibliography,
lengthcheck
]{revtex4-1}
\usepackage{graphicx}

\def\vec#1{\mbox{\boldmath $#1$}}
\newcommand{\vd}{\mathcal{D}}

\begin{document}

\title{Lifetime of dynamical heterogeneity in a highly supercooled liquid} 

\author{Hideyuki Mizuno}
 \email{h-mizuno@cheme.kyoto-u.ac.jp}
\author{Ryoichi Yamamoto}
 \email{ryoichi@cheme.kyoto-u.ac.jp}
\affiliation{Department of Chemical Engineering, Kyoto University, Kyoto 615-8510, Japan}

\date{\today}

\begin{abstract}
We numerically examine dynamical heterogeneity in a highly supercooled three-dimensional liquid
via molecular-dynamics simulations.
To define the local dynamics, we consider two time intervals, $\tau_\alpha$ and $\tau_{\text{ngp}}$.
$\tau_\alpha$ is the $\alpha$ relaxation time, and $\tau_{\text{ngp}}$ is the time at which non-Gaussian parameter of the van Hove self-correlation function is maximized.
We determine the lifetimes of the heterogeneous dynamics in these two different time intervals,
$\tau_{\text{hetero}}(\tau_\alpha)$ and $\tau_{\text{hetero}}(\tau_{\text{ngp}})$,
by calculating the time correlation function of the particle dynamics, i.e., the four-point correlation function.
We find that the difference between $\tau_{\text{hetero}}(\tau_\alpha)$ and $\tau_{\text{hetero}}(\tau_{\text{ngp}})$ increases with decreasing temperature.
At low temperatures, $\tau_{\text{hetero}}(\tau_\alpha)$ is considerably larger than $\tau_{\alpha}$,
while $\tau_{\text{hetero}}(\tau_{\text{ngp}})$ remains comparable to $\tau_{\alpha}$.
Thus, the lifetime of the heterogeneous dynamics depends strongly on the time interval.
\end{abstract}


\maketitle 

One of the long-unresolved problems in material science is the glass transition \cite{ediger_1996} \cite{Debenedetti_2001}.
In spite of the extremely widespread use of glass in industry,
the formation process and dynamical properties of this material are still poorly understood.
Numerous studies have attempted to explain the fundamental mechanisms of the slowing of the dynamics 
observed in fragile glass (i.e., the sharp increase in viscosity in the vicinity of the glass transition).
However, the physical mechanisms behind the glass transition have not been successfully identified.

Recently, \textit{dynamical heterogeneities} in glass-forming liquids have attracted much attention.
In a system displaying dynamical heterogeneity, the dynamic characteristics
(i.e., particle displacements and local structural relaxations) are non-uniformly distributed throughout space.
Dynamical heterogeneities have been detected and visualized through
simulations of soft-sphere systems \cite{muranaka_1994} \cite{hurley_1995} \cite{yamamoto1_1998} \cite{yamamoto_1998} \cite{perera_1999} \cite{cooper_2004},
hard-sphere systems \cite{doliwa_2002},
Lennard-Jones (LJ) systems \cite{donati_1998},
and experiments \cite{kegel_2000} \cite{weeks_2000}.
Insight into the mechanisms of dynamical heterogeneities will lead to a better understanding of the slowing of the dynamics near the glass transition.

Conventional two-point density correlation functions
are not informative when applied to the investigation of dynamical heterogeneities. 
We need to examine the correlation of the particle dynamics, not just snapshots.
We can quantify the correlation length $\xi$ of the heterogeneous dynamics by calculating the four-point correlation functions, which correspond to the static structure factor of the particle dynamics.
Several simulations \cite{yamamoto1_1998} \cite{lacevic_2003} \cite{toninelli_2005} \cite{stein_2008},
experiments \cite{ediger_2000} \cite{berthier_2005},
and mode-coupling theory \cite{biroli_2006}
have estimated $\xi$ in terms of the four-point correlation functions
and revealed that $\xi$ increases with decreasing temperature.
In addition, we can quantify the lifetime $\tau_{\text{hetero}}$ of the heterogeneous dynamics by
employing the multiple time extension of the four-point correlation functions
(i.e., the multi-time correlation functions),
which correspond to the time correlation functions of the particle dynamics.
$\tau_{\text{hetero}}$ has been measured in terms of the multi-time correlation functions by
simulations \cite{yamamoto1_1998} \cite{yamamoto_1998} \cite{flenner_2004} \cite{kim_2009}
and experiments \cite{ediger_2000} \cite{wang_1999} \cite{wang_2000}.
It was reported that $\tau_{\text{hetero}}$ increases dramatically with decreasing temperature and
can become greater than the $\alpha$ relaxation time near the glass transition.

In 2009, Kim et al.\cite{kim_2009} investigated the correlations
between the heterogeneous dynamics at various time intervals.
They calculated the sum of the time correlation functions
for the heterogeneous dynamics at these time intervals 
and determined the lifetime of the heterogeneous dynamics as a characteristic time
at which the sum of correlation functions decays.

In this letter, we demonstrate via molecular-dynamics (MD) simulations that
the lifetime of the heterogeneous dynamics depends strongly on the time interval.
To define the local dynamics, we consider two time intervals: the $\alpha$ relaxation time, $\tau_\alpha$, and the time $\tau_{\text{ngp}}$ at which the 
non-Gaussian parameter of the van Hove self-correlation function is maximized.
We estimate the lifetimes of the heterogeneous dynamics in these two different time intervals 
,$\tau_{\text{hetero}}(\tau_\alpha)$ and $\tau_{\text{hetero}}(\tau_{\text{ngp}})$,
by calculating the time correlation function of the particle dynamics.
Finally, we compare the two lifetimes.

\begin{figure}[t]
\begin{center}
\includegraphics[scale=0.39, angle=0]{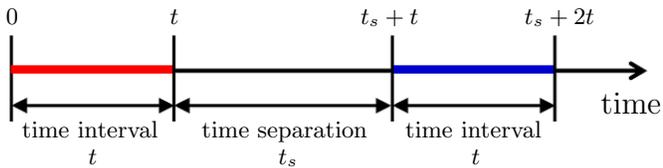}
\put(-242,-6){time interval}
\put(-217,-16){$t$}
\put(-174,-6){time separation}
\put(-145,-16){$t_s$}
\put(-97,-6){time interval}
\put(-72,-16){$t$}
\put(-248,37){$0$}
\put(-186,37){$t$}
\put(-114,37){$t_s+t$}
\put(-52,37){$t_s+2t$}
\put(-23,3){{\large time}}
\end{center}
\vspace*{-3mm}
\caption{Schematic illustration of two time intervals and their time separation.}
\label{ts}
\end{figure}

The conventional two-point correlation function $F({k},t)$
represents the correlation of the local fluctuations $\delta n(\vec{k},t)$ in some order parameter, such as the particle density.
$\delta n(\vec{k},t)$ is the Fourier component \vec{k} of the fluctuations at the time $t$, and
$F({k},t) = \langle \delta n(\vec{k},t) \delta n(-\vec{k},0) \rangle$, $k=\left| \vec{k} \right|$.
The two-point correlation function can
describe the particle dynamics in the time interval $[0,t]$, averaged over the initial time and space.
As the time interval $t$ increases, $F({k},t)$ decays in the stretched exponential form,
\begin{equation}
\frac{F({k},t)}{F({k},0)} \sim \exp \left( -\left( \frac{t}{\tau({k})}\right)^\beta \right),
\end{equation}
where $\tau({k})$ is the relaxation time of the two-point correlation function, which
represents the characteristic timescale of the averaged particle dynamics.
To examine the lifetime of spatially heterogeneous dynamics,
we have to calculate the time correlation function of the local fluctuations $\delta Q_{{k}}(\vec{q},t_0,t)$ in the particle dynamics.
$\delta Q_{{k}}(\vec{q},t_0,t)$ is the Fourier component \vec{q} of the fluctuations in the particle dynamics associated with a microscopic
wavenumber ${k}$ in the time interval $[t_0,t_0+t]$.
$F({k},t)$ is equal to $Q_{{k}}(\vec{q},t_0,t)$ averaged over the initial time $t_0$ and space, i.e., $F({k},t) \sim \langle Q_{{k}}(\vec{q},t_0,t) \rangle$.
The time correlation function defined by
\begin{equation}
\begin{aligned}
F_{4,{k}}({q},t_s,t)=
\langle \delta Q_{{k}}(\vec{q},t_s+t,t) \delta Q_{{k}}(-\vec{q},0,t) \rangle,
\end{aligned} \label{F4}
\end{equation}
represents the correlation of particle dynamics between the two time intervals $[0,t]$ and $[t_s+t,t_s+2t]$.
$t_s$ is the time separation between the two time intervals $[0,t]$ and $[t_s+t,t_s+2t]$, as is
schematically illustrated in Fig.\ref{ts}.
$F_{4,{k}}({q},t_s,t)$
is the multiple time extension of the four-point correlation function \cite{kim_2009}.
As the time separation $t_s$ increases, $F_{4,{k}}({q},t_s,t)$ with fixed $t$ decays
in the stretched exponential form,
\begin{equation}
\begin{aligned}
\frac{F_{4,{k}}(q,t_s,t)}{F_{4,{k}}({q},0,t)}
\sim \exp\left( - \left( \frac{t_s}{\tau_{4,{k}}({q},t)} \right)^c \right),
\end{aligned} \label{F42}
\end{equation}
where $\tau_{4,{k}}({q},t)$ is the relaxation time of the correlation of the particle dynamics.
We determined the lifetime of the heterogeneous dynamics $\tau_{\text{hetero}}(t)$
as $\tau_{4,{k}}({q},t)$ at $q=0.38$, the smallest wavenumber in our simulation.

To calculate the time correlation function of the particle dynamics,
we performed MD simulations in three dimensions on binary mixtures of
two different atomic species, 1 and 2,
with $N_1=N_2=5,000$ particles and a cube of constant volume $V$ as the basic cell,
surrounded by periodic boundary image cells.
The particles interact via the soft-sphere potentials\ 
$v_{a b}(r)= \epsilon (\sigma_{a b}/r)^{12}$,
where r is the distance between two particles, $\sigma_{a b} = (\sigma_a + \sigma_b)/2$, and
$a,b \in 1,2$. The interaction was truncated at $r=3 \sigma_{a b}$.
In the present letter, the following
dimensionless units were used: length, $\sigma_1$; temperature, $\epsilon/k_B$; and time, $\tau_0 = (m_1 \sigma_1^2/\epsilon)$.
The mass ratio was $m_2/m_1=2$, and the diameter ratio was $\sigma_2/\sigma_1=1.2$. This diameter ratio avoided system crystallization 
and ensured that an amorphous supercooled state formed at low temperatures \cite{miyagawa_1991}.
The particle density was fixed at the high value of $\rho = (N_1 + N_2)\sigma_1^3/V = 0.8$. The system length was
$L=V^{1/3}=23.2\sigma_1$.
Simulations were carried out at $T=0.772,\ 0.473,\ 0.352,\ 0.306,\ 0.289,\ 0.267$, and $0.253$.
Note that the freezing point of the corresponding one-component model is around $T=0.772\ (\Gamma_{\text{eff}}=1.15)$
\cite{miyagawa_1991}.
Here, $\Gamma_{\text{eff}}$ is the effective density, a single parameter characterizing this model.
At $T=0.253\ (\Gamma_{\text{eff}}=1.52)$, the system is in a highly supercooled state.
We used the leapfrog algorithm with time steps of $0.005\tau$
when integrating the Newtonian equation of motion.
Very long annealing times ($3 \times 10^6$ for $T=0.253$) were chosen.
No appreciable aging effect was detected in various quantities, including the pressure or 
the density correlation function.

\begin{figure}[t]
\begin{center}
\includegraphics[scale=0.31, angle=270]{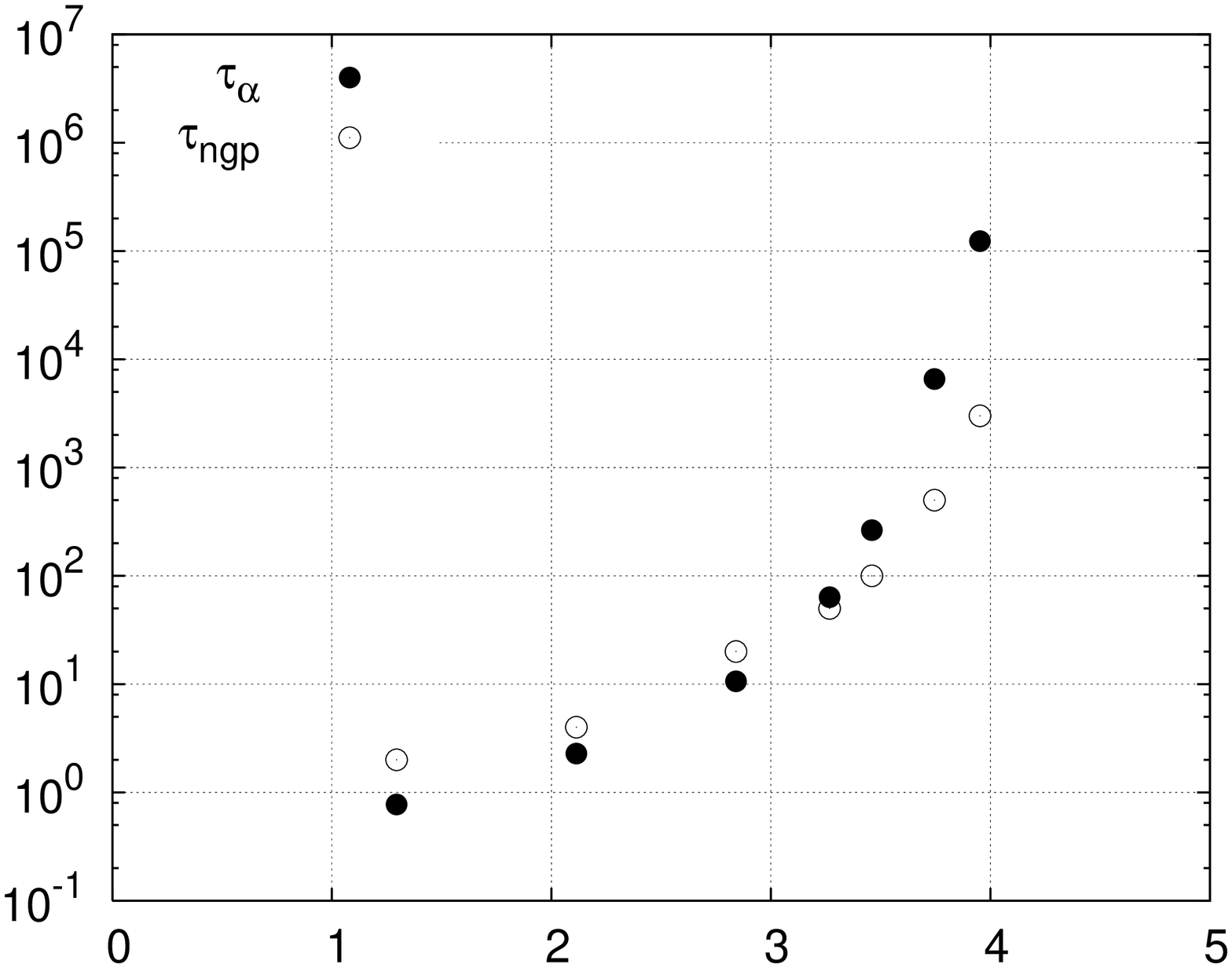}
\put(-100,-164){$1/T$}
\put(-209,-80){\rotatebox{90}{$\tau_\alpha,\ \tau_{\text{ngp}}$}}
\end{center}
\vspace*{-3mm}
\caption{$\tau_\alpha$,\ $\tau_{\text{ngp}}$ versus inverse temperature $1/T$. We 
use these two time intervals to define the local dynamics.}
\label{taua}
\end{figure}

\begin{figure*}
\begin{minipage}{0.42\textwidth}
\begin{center}
\includegraphics[scale=0.40]{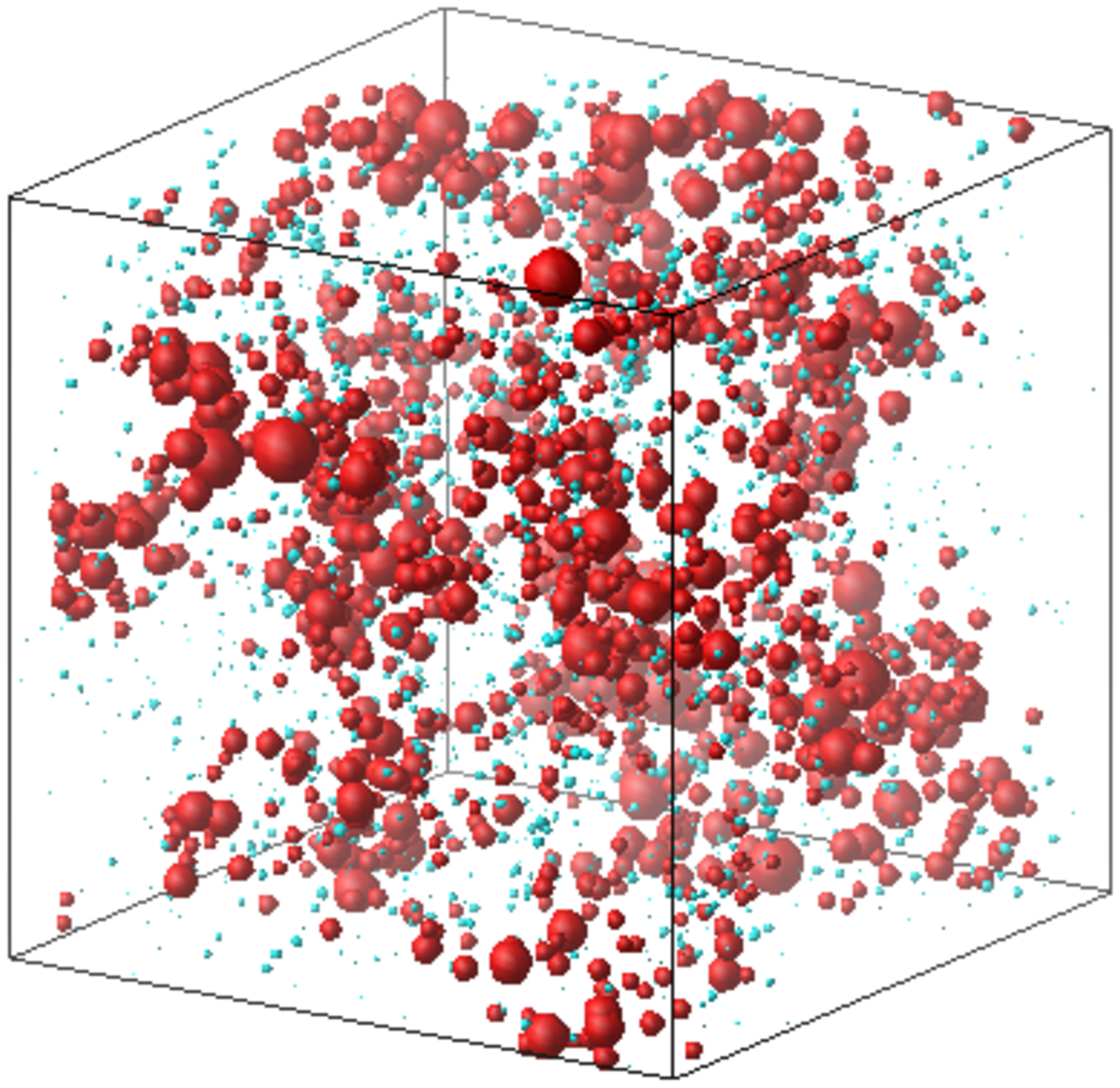}
\\
\vspace*{-5mm}
(a)
\end{center}
\end{minipage}
\begin{minipage}{0.42\textwidth}
\begin{center}
\includegraphics[scale=0.40]{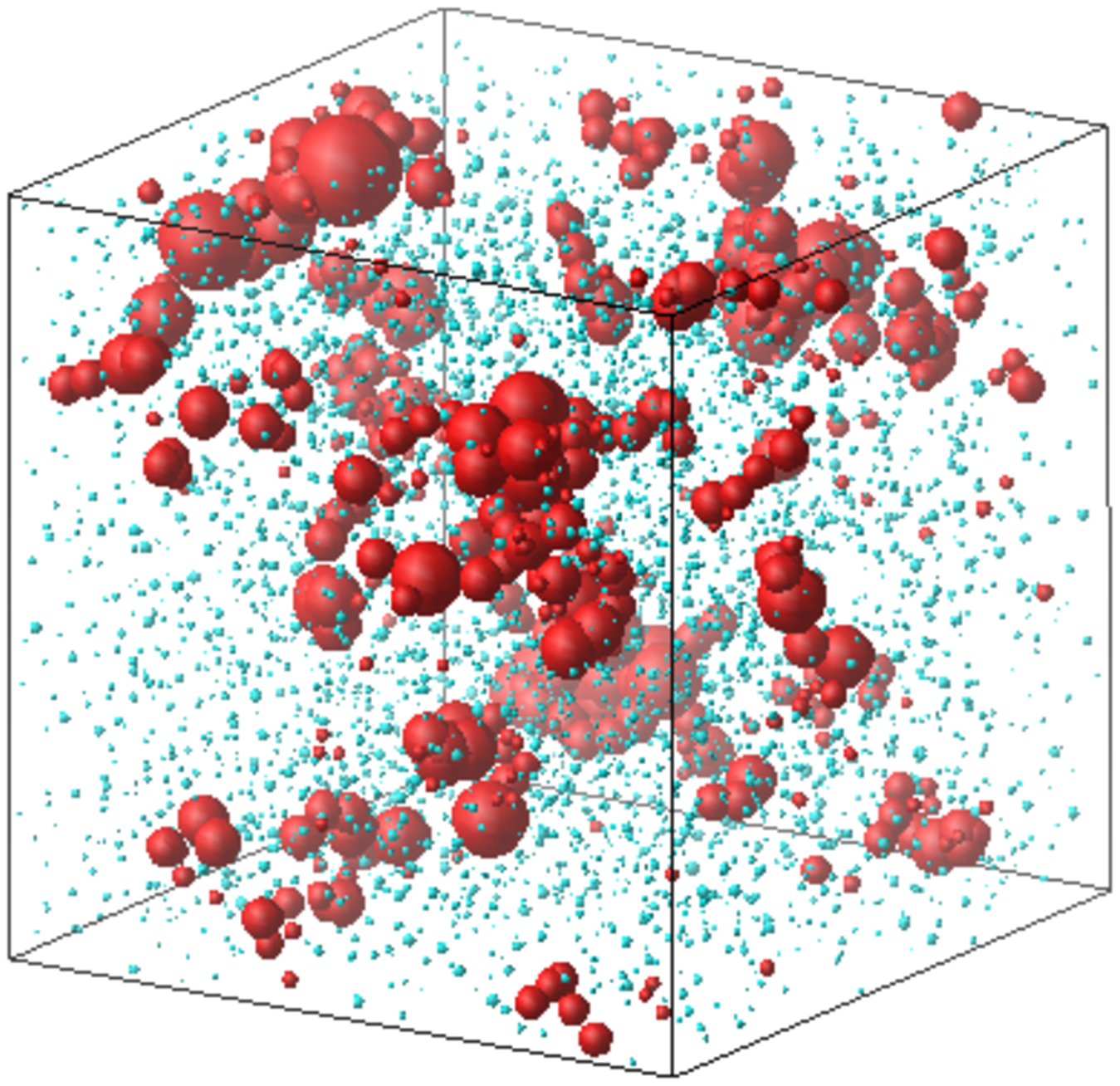}
\\
\vspace*{-5mm}
(b)
\end{center}
\end{minipage}
\caption{Visualization of the heterogeneous dynamics of particle species 1.
The temperature is $0.253$.
The time interval is $[t_0, t_0+\tau_\alpha]$ in (a), $[t_0, t_0+\tau_{\text{ngp}}]$ in (b).
The radii of the spheres are 
$\left| \Delta \vec{r}_j(t_0,t)  \right|^2/ {\langle [\Delta \vec{r}_j(t_0,t)]^2 \rangle }$,
and the centers are at $\frac{1}{2}[\vec{r}_j(t_0) + \vec{r}_j(t_0+t)]$.
The red spheres and blue spheres represent $\left| \Delta \vec{r}_j(t_0,t)  \right|^2/ {\langle [\Delta \vec{r}_j(t_0,t)]^2 \rangle }\ge 1$ and
$\left| \Delta \vec{r}_j(t_0,t)  \right|^2/ {\langle [\Delta \vec{r}_j(t_0,t)]^2 \rangle }<1$, respectively.
}
\label{hetero}
\end{figure*}

To define the local dynamics, we consider the two time intervals, $\tau_\alpha$ and $\tau_{\text{ngp}}$.
$\tau_\alpha$ is the $\alpha$ relaxation time defined by $F_s(k_m,\tau_{\alpha}) = e^{-1}$,
where $F_s(k,t)$ is the self-part of the density time correlation function for particle species 1, and
$k_m = 2\pi$ is the first peak wavenumber of the static structure factor.
$\tau_{\text{ngp}}$ is the time at which the non-Gaussian parameter \cite{rahman_1964}
of the van Hove self-correlation function is maximized.
In Fig.\ref{taua}, we show $\tau_\alpha$ and $\tau_{\text{ngp}}$ as function of the inverse temperature $1/T$.
$\tau_\alpha \simeq \tau_{\text{ngp}}$ at $T=0.306$, and
$\tau_\alpha$ grows exponentially larger than $\tau_{\text{ngp}}$ with decreasing temperature at $T<0.306$.
This trend agrees with other simulation results of Lennard-Jones (LJ) systems \cite{kob_1995_1} \cite{kob_1995_2}.
Next, we visualize the heterogeneous dynamics of $\tau_\alpha$ and $\tau_{\text{ngp}}$
in the same manner presented in Ref.\cite{yamamoto_1998}.
We calculate the displacement of each particle of species 1 in
the time interval $[t_0, t_0 + t]$,
$\Delta \vec{r}_j(t_0,t) =\vec{r}_j(t_0+t)-\vec{r}_j(t_0) \ (j=1,2,...,N_1)$.
In Fig.\ref{hetero}, particles are drawn as spheres with radii
\begin{equation}
a_j^2(t_0,t) = \frac {\left| \Delta \vec{r}_j(t_0,t)  \right|^2}{\langle [\Delta \vec{r}_j(t_0,t)]^2 \rangle}
\end{equation}
located at $\vec{R}_j(t_0,t) = \frac{1}{2}[\vec{r}_j(t_0) + \vec{r}_j(t_0+t)]$
in both time intervals: $[t_0, t_0+\tau_\alpha]$ ($t=\tau_\alpha$) in \ref{hetero}(a),\ 
$[t_0,t_0+\tau_{\text{ngp}}]$ ($t=\tau_{\text{ngp}}$) in \ref{hetero}(b).
The temperature is $0.253$.
$a_j^2(t_0,t)\ge 1$ ($a_j^2(t_0,t)<1$) means that the particle $j$ moves more (less) than the mean value of the single-particle displacement.
In Fig.\ref{hetero}, the red (blue) spheres represent $a_j^2(t_0,t)\ge 1$ ($a_j^2(t_0,t)<1$).
We can see the large-scale heterogeneities in both $\tau_\alpha$ and $\tau_{\text{ngp}}$.

\begin{figure}[b]
\begin{minipage}[h]{0.49\textwidth}
\begin{center}
\includegraphics[scale=0.31, angle=270]{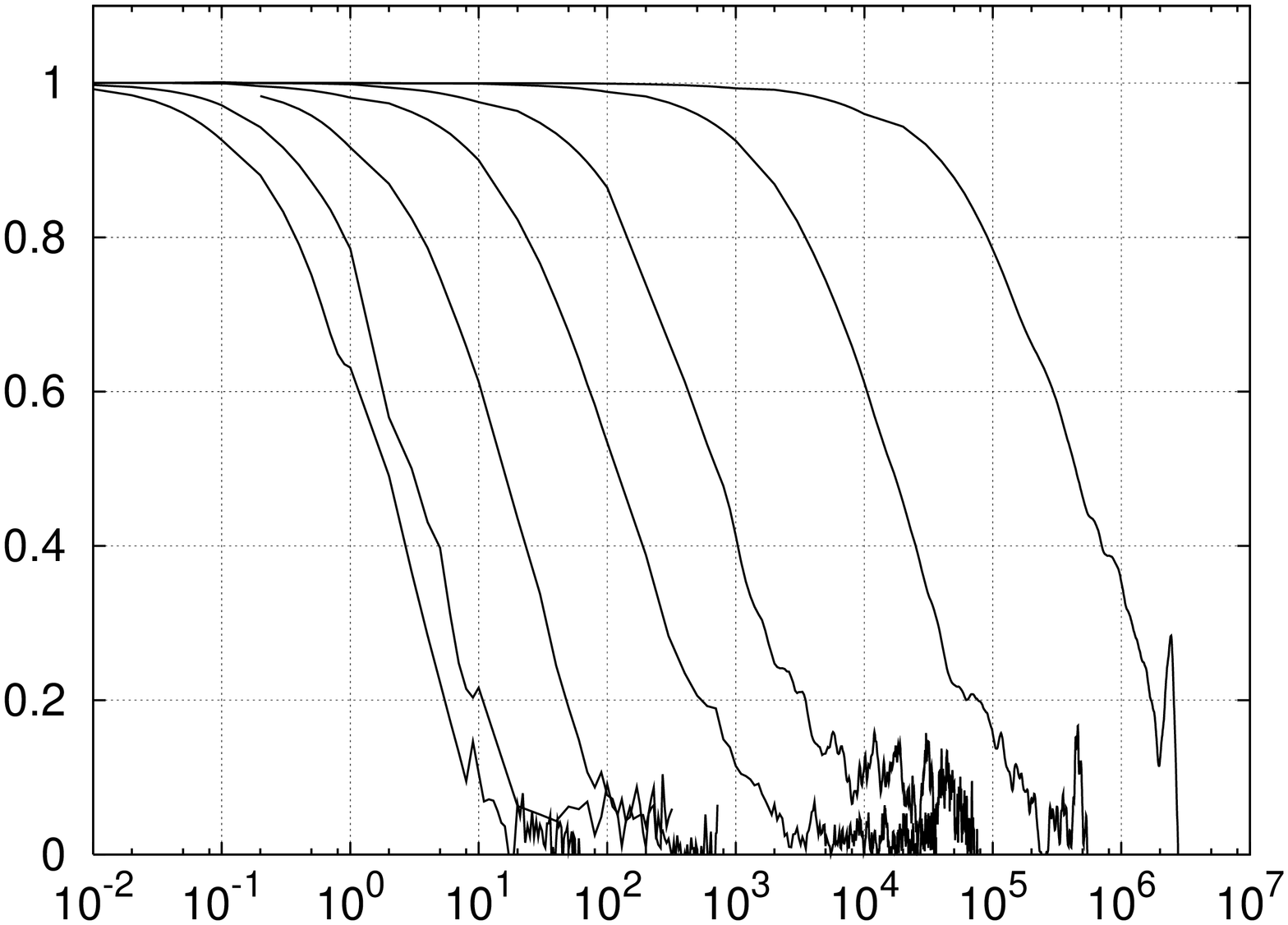}
\put(-100,-162){$t_s$}
\put(-223,-110){\rotatebox{90}{${S_{\vd}(q,t_s,t)}/{S_{\vd}(q,0,t)}$}}
\put(-185,-110){$T=0.772$}
\put(-55,-35){$T=0.253$}
\end{center}
\end{minipage}
\vspace*{-1mm}
\caption{
The time decay of $S_{\vd}(q,t_s,t)$ at $t=\tau_\alpha$, $q=0.38$ for $T=0.772$ - $0.253$.
$q=0.38$ is the smallest wavenumber in our simulation.
Temperature decreases going right.
}
\label{decay}
\end{figure}

\begin{figure}[b]
\begin{center}
\includegraphics[scale=0.32, angle=270]{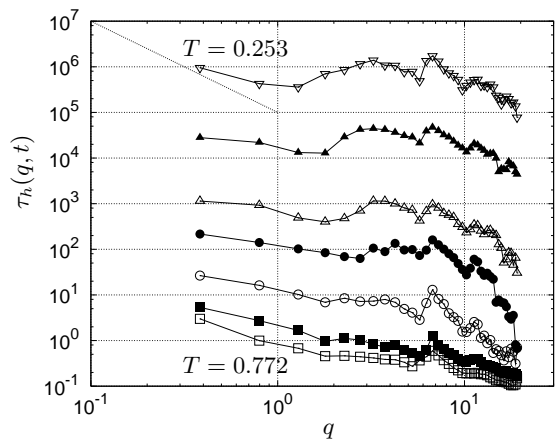}
\put(-100,-165){$q$}
\put(-218,-80){\rotatebox{90}{$\tau_{h}(q,t)$}}
\put(-153,-142){$T=0.772$}
\put(-153,-22){$T=0.253$}
\end{center}
\vspace*{-4mm}
\caption{The wavenumber dependence of $\tau_{h}(q,t)$ at $t=\tau_\alpha$ for $T=0.772-0.253$. 
Temperature decreases going up.
The dotted line is $\tau_h(q,\tau_\alpha) \sim q^{-2}$}
\label{tauhq}
\end{figure}

We can quantify the lifetimes of the heterogeneous dynamics in both time intervals.
We consider the local fluctuations in the particle dynamics defined by
\begin{equation}
\delta \vd(\vec{q},t_0,t) = \sum_{j=1}^{N_1} (a_j^2(t_0,t)-1) \exp[-i\vec{q} \cdot \vec{R}_j(t_0,t) ],
\end{equation}
which is equal to the fluctuations in the \textit{diffusivity} density defined in ref \cite{yamamoto_1998} and
represents the local fluctuations in the particle dynamics in the time interval $[t_0,t_0+t]$.
We use $\delta \vd(\vec{q},t_0,t)$ as $\delta Q_{{k}}(\vec{q},t_0,t)$ in Eq.(\ref{F4}), and
the time correlation function defined by
\begin{equation}
S_{\vd}(q,t_s,t) = \langle \delta \vd(\vec{q},t_s+t,t) \delta \vd(-\vec{q},0,t) \rangle,
\end{equation}
corresponds to $F_{4,{k}}({q},t_s,t)$.
$S_{\vd}(q,t_s,t)$ represents the correlation of the particle dynamics between two time intervals $[0,t]$ and $[t_s+t,t_s+2t]$
(see Fig.\ref{ts}).
Thus, we can estimate the lifetime of the heterogeneous dynamics
by examining the time decay of $S_{\vd}(q,t_s,t)$.
As the time separation $t_s$ increases, $S_{\vd}(q,t_s,t)$ with fixed $t$ decays
in the stretched exponential form,
\begin{equation}
\begin{aligned}
\frac{S_{\vd}(q,t_s,t)}{S_{\vd}(q,0,t)}
\sim \exp\left( - \left( \frac{t_s}{\tau_{h}(q,t)} \right)^c \right),
\end{aligned}
\end{equation}
where $\tau_{h}(q,t)$ is the wavenumber-dependent heterogeneous dynamics lifetime, which
corresponds to $\tau_{4,{k}}({q},t)$ in Eq.(\ref{F42}).
Figure \ref{decay} shows the time decay of $S_{\vd}(q,t_s,t)$ at $t=\tau_\alpha$, $q=0.38$ for various temperatures.
$q=0.38$ is the smallest wavenumber in our simulation.
In Fig.\ref{tauhq}, we show the wavenumber dependence of $\tau_{h}(q,t)$ at $t=\tau_\alpha$. 
$\tau_{h}(q,\tau_\alpha)$ depends on $q$ more weakly than the $q$-dependent relaxation time of the two-point density correlation functions
and dramatically increases with decreasing temperature in a wide region of $q$ ($q=0.38-19$).
Furthermore, we can see that $\tau_{h}(q,\tau_\alpha)$ approaches $\tau_{h}(q,\tau_\alpha) \sim q^{-2}$ at small wavenumbers.
This suggests that the heterogeneous dynamics migrate in space with a diffusion-like mechanism.
These results of $\tau_{h}(q,\tau_\alpha)$ are qualitatively the same as those of $\tau_{h}(q,\tau_{\text{ngp}})$.

\begin{figure}
\begin{center}
\includegraphics[scale=0.32, angle=270]{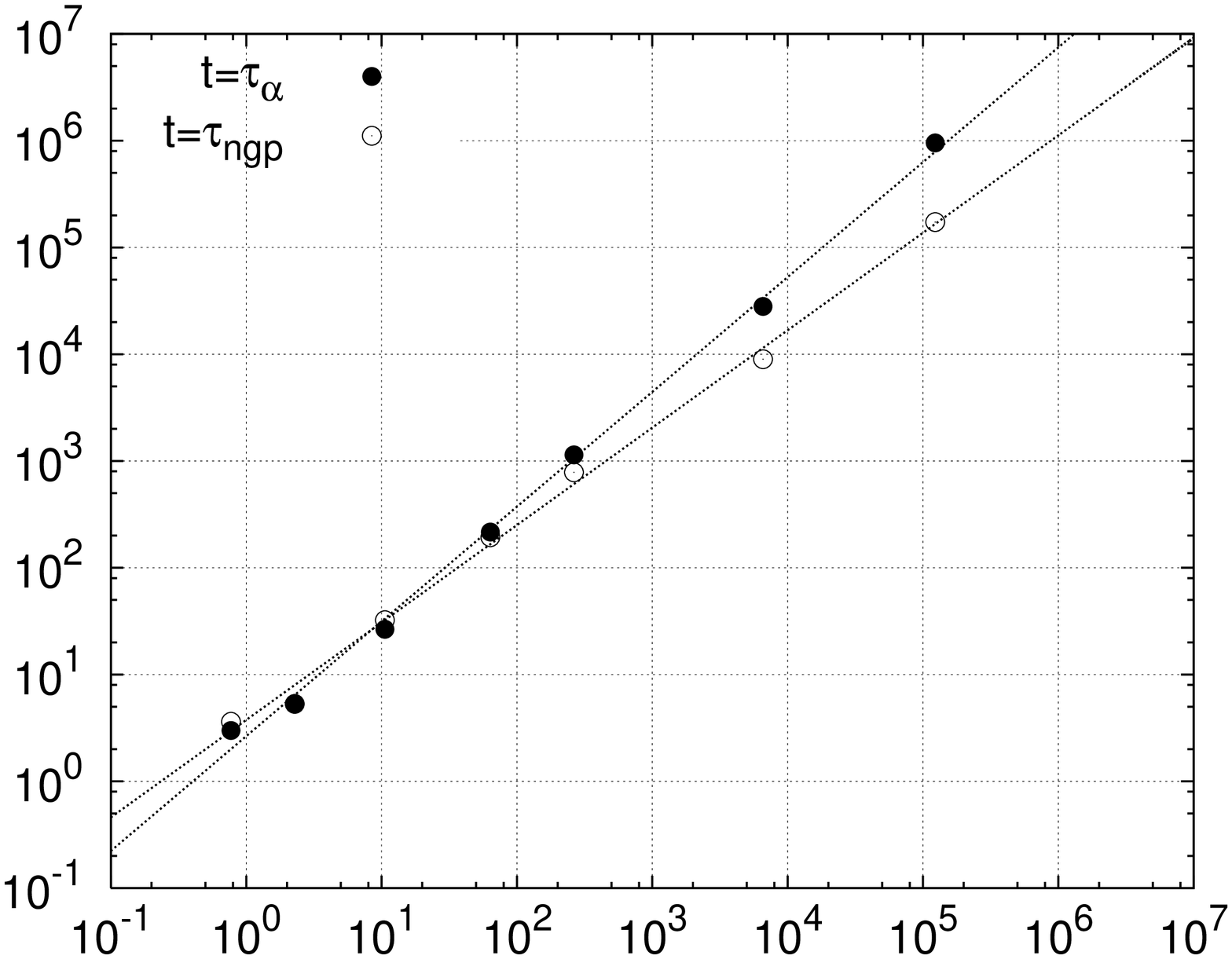}
\put(-100,-167){$\tau_\alpha$}
\put(-217,-80){\rotatebox{90}{$\tau_{\text{hetero}}(t)$}}
\put(-145,-44){$\tau_{\text{hetero}}=2.65\tau_{\alpha}^{1.08}$}
\put(-91,-73){$\tau_{\text{hetero}}=3.75\tau_{\alpha}^{0.91}$}
\end{center}
\vspace*{-4mm}
\caption{The lifetime $\tau_{\text{hetero}}(t)$ 
for $t=\tau_\alpha$,\ $\tau_{\text{ngp}}$ versus $\tau_\alpha$. 
The line $\tau_{\text{hetero}}=2.65\tau_{\alpha}^{1.08}$ is fitted
for $t=\tau_\alpha$, while 
the line $\tau_{\text{hetero}}=3.75\tau_{\alpha}^{0.91}$ is fitted
for $t=\tau_{\text{ngp}}$.
}
\label{tauatauh}
\end{figure}

We determined the lifetime of the heterogeneous dynamics $\tau_{\text{hetero}}(t)$ as $\tau_{h}(q,t)$ at $q=0.38$, which is
the time separation $t_s$ at which $S_{\vd}(q,t_s,t)/S_{\vd}(q,0,t)$ at $q=0.38$ equals $e^{-1}$ in Fig.\ref{decay}.
$\tau_{\text{hetero}}(t)$ increases dramatically with decreasing temperature.
We plot $\tau_{\text{hetero}}(t)$ versus $\tau_\alpha$ in Fig.\ref{tauatauh}, which shows that
$\tau_{\text{hetero}}(\tau_\alpha) \sim \tau_{\alpha}^{1.08}$ and
$\tau_{\text{hetero}}(\tau_{\text{ngp}}) \sim \tau_{\alpha}^{0.91}$.
The difference between
$\tau_{\text{hetero}}(\tau_\alpha)$ and $\tau_{\text{hetero}}(\tau_{\text{ngp}})$
increases with decreasing temperature.
At $T=0.253$, $\tau_{\text{hetero}}(\tau_\alpha) \simeq 7.8\tau_{\alpha}$, and
$\tau_{\text{hetero}}(\tau_{\text{ngp}}) \simeq 1.4\tau_{\alpha}$.
Therefore,
$\tau_{\text{hetero}}(\tau_\alpha)$ is considerably larger than $\tau_{\alpha}$,
while $\tau_{\text{hetero}}(\tau_{\text{ngp}})$ is comparable to $\tau_{\alpha}$.
The existence of a slower timescale in the heterogeneous dynamics
is consistent with Ref.\cite{kim_2009}.

Finally, we examine the finite-size effect. 
To this end, we performed MD simulations using a larger system with $N_1=N_2=50,000$ and $L=50 \sigma_1$,
and compared our results with those of a larger system.
No finite-size effect was detected
in quantities such as $\tau_{\alpha},\ \tau_{\text{ngp}}$, or $\tau_{\text{hetero}}$.

In summary, we have investigated the heterogeneous dynamics in two different time intervals,
$\tau_\alpha$ and $\tau_{\text{ngp}}$.
We quantified the lifetimes of the heterogeneous dynamics in these two intervals,
$\tau_{\text{hetero}}(\tau_\alpha)$ and $\tau_{\text{hetero}}(\tau_{\text{ngp}})$,
by calculating the time correlation function of the particle dynamics.
We found that the difference between
$\tau_{\text{hetero}}(\tau_\alpha)$ and $\tau_{\text{hetero}}(\tau_{\text{ngp}})$
increases with decreasing temperature.
At low temperatures, $\tau_{\text{hetero}}(\tau_\alpha)$ is considerably larger than $\tau_{\alpha}$,
while $\tau_{\text{hetero}}(\tau_{\text{ngp}})$ remains comparable to $\tau_{\alpha}$.
Thus, we can conclude that the lifetime of the heterogeneous dynamics depends strongly on the time interval.
We also have examined the finite-size effect. No finite-size effect was detected in our study. 
\bibliography{Letter}
\end{document}